# ChinMotion Rapidly Enables 3D Computer Interaction after Tetraplegia

Ferran Galán[1,2*], Stuart N. Baker[2] and Monica A. Perez[1]

[1]Department of Neurological Surgery, The Miami Project to Cure Paralysis, University of Miami. [2]Institute of Neuroscience, Newcastle University.

Correspondence to:
Ferran Galán PhD.
Henry Wellcome Building
The Medical School, Framlington Place
Newcastle upon Tyne, NE2 4HH, UK
Telephone: +44 (0) 191 208 3271
Fax: +44(0) 191 208 5227
ferran.galan@ncl.ac.uk

# Abstract

Individuals with severe paralysis require hands-free interfaces to control assistive devices that can improve their quality of life. We present ChinMotion, an interface that noninvasively harnesses preserved chin, lip and tongue sensorimotor function after tetraplegia to convey intuitive control commands. After two hours of practice, ChinMotion enables superior point-and-click performance over existing interfaces and it facilitates accurate 3D control of a virtual robotic arm.

# Main

Humans with spinal cord injury[1-8], stroke[5,9-11], and other disorders[10,12-16] producing limited volitional movement have benefited from hands-free machine interfaces to interact with technology that can enhance their quality of life. On one hand, brain-machine interfaces[17] which directly decode movement intentions from brain activity have enabled multidimensional control of prosthetic limbs[7,9,12,16], but current prototypes lack kinaesthetic feedback and require extensive training before users learn the artificial mapping between brain activity and prosthetic behaviour. On the other hand, body-machine interfaces[18] create kinaesthetic maps between body motions and their functional and sensory consequences by translating preserved sensorimotor actions, such as upper-body motions or muscle contractions, into inputs that have enabled control of computer pointers and wheelchairs in a more intuitive manner[2-6,8,14,15]. However, kinaesthetic maps rapidly enabling accurate multidimensional prosthetic control in individuals with paralysis remain an unmet goal.

We developed an open-source body-machine interface (ChinMotion; **Supplementary Software 1-2; Supplementary Schematics; Supplementary Step-by-Step Instructions**) that translates chin, lip and tongue motion into intuitive 3D control commands with minimal training. We tested ChinMotion in humans with high cervical SCI who experienced severe paralysis of upper and lower-limb muscles (tetraplegia). Muscles of the face and tongue innervated from cranial nerves are typically not damaged after a high cervical injury; they therefore provide an excellent source of preserved somatosensory feedback to control machine interfaces. ChinMotion is a wearable pointing device that integrates three main components: (1) a triple axis accelerometer placed below the lower lip which can be tilted by lateral, protrusive and retrusive movements of the chin and lips, (2) a conductive rubber stretchable cord sensor following the contour of the face which can be stretched by protrusive movements of the tongue, and (3) a microcontroller that translates in real time inertial and resistance measurements from (1) and (2) into (XY) and (“click”/Z) controls respectively (**Fig. 1a; Fig. 2a, d-e**).

To test ChinMotion, we examined point-and-click performance of eight individuals with severe tetraplegia (**Supplementary Table 1**) and eight age-matched controls participating in a centre-out-centre reach-and-click task in a single session. Participants first moved the pointer from the centre of the screen to select a randomly-appearing peripheral target and then moved back to select the centre (Methods; **Fig. 2a; Supplementary Video 1**). Uninjured controls also performed the task using a keyboard mounted pointing stick (**Fig. 1b**) which was used as a gold standard for comparison with ChinMotion. Participants had no previous ChinMotion or pointing stick experience. To estimate point-and-click performance, we used Fitts’ law, throughput, and error rates following the ISO9241-9 standard recommendations[19] (see Methods). Such a framework allowed for comparison with mainstream hand-operated pointing devices[19], and hands-free body- and brain- machine interfaces with the highest performance reported to date evaluating the same metrics: the Tongue Drive System[4] (TDS) and the BrainGate neuroprosthetic system[11].

As predicted by Fitt’s law, we found that all participants completed the task dedicating longer target selection times as difficulty increased (**Supplementary Table 2; Supplementary Figure 1**). The throughput of uninjured controls (0.54±0.05 bits/s, bits per second) and individuals with tetraplegia (0.55±0.07 bits/s) using ChinMotion were similar, suggesting comparable target selection times across the range of target difficulties tested in a single session ($p > 0.05$, Wilcoxon rank – sum test; **Fig. 2b**). Notably, the error rates of uninjured controls (18.4±3.8%) and individuals with tetraplegia (22.0±5.6%) were also comparable in a single session ($p > 0.05$, Wilcoxon rank – sum test; **Fig. 2c**), suggesting similar accuracy across groups. Performance was worse than for uninjured controls using the pointing stick (**Supplementary Table 2; Fig. 2b, c**) consistent with previous findings

**(Supplementary Table 3)**. Importantly, our results indicate that point-and-click performance using ChinMotion is superior than that achieved using state-of-the art high-performance hands-free body- and brain-machine interfaces[4,11] (Methods). Point-and-click performance of individuals with tetraplegia using ChinMotion in the first session was about 2 times faster and 3 times more accurate than using the Tongue-Drive System[4] (TDS; Methods; **Supplementary Table 4**) and it was 1.5 times faster and 2 times more accurate than using the BrainGate neuroprosthetic system[11] after 1000 days (Methods; **Supplementary Table 6**).

We examined ChinMotion 3D control of a virtual robotic arm in a centre-out-centre reach-and-hold task within the same experimental session (Methods; **Fig. 2d, e; Supplementary Video 2**). Here, participants moved the arm from a starting location and held its endpoint within the volume of an instructed target for 1 second. They then moved back, holding at the starting location for another second. Here, we found that all participants autonomously controlled the arm with 100% success rates. Furthermore, uninjured controls and individuals with tetraplegia completed the task at comparable speeds, requiring average trial completion times of 17.5±2.6 and 22.5±3.5 seconds, respectively ($p > 0.05,$ Wilcoxon $\text{rank} - \text{sum}$ test; **Fig. 2f**).

Altogether, our results demonstrate that preserved sensorimotor pathways in the face offer a reproducible avenue towards the restoration of rapid and precise 3D computer interaction after severe paralysis. Unlike current hands-free interfaces, ChinMotion exploits preserved sensorimotor pathways rapidly to establish intuitive kinaesthetic maps between motor commands and end effectors, which allows precise control with minimal training. Enabled by ChinMotion, individuals with high tetraplegia participating in our study achieved levels of point-and-click performance only reached before with current high-performing interfaces[4,11] after several days of training. Furthermore, to our knowledge, ChinMotion is the first hands-free machine interface enabling precise 3D control of a virtual robotic arm for individuals with tetraplegia after only two hours of practice, demonstrating the feasibility of our approach to facilitating intuitive control of prosthetic devices for restoring limb function. ChinMotion is released as an open-source design for public use (**Supplementary Software 1-2; Supplementary Schematics; Supplementary Step-by-Step Instructions)** with the aim of stimulating the innovation of body-machine interfaces for research, educational, and development purposes. ChinMotion can be easily replicated, modified and improved, serving as a valuable resource for the development of affordable high-performance interfaces capable of enabling multidimensional prosthetic control. Coupled with advances in wearable flexible electronics and rapid-prototyping methods, we hope that this approach will pave the way for a new generation of mainstream wearable control interfaces. Such devices are likely to be the most plausible option to restore independent function for the majority of individuals with paralysis.

## Methods

**Participants.** Eight individuals with chronic (≥ 1 year) cervical injury (C2-C4; mean age = 36.5±2.4 years, 2 female) classified by the International Standards for Neurological Classification of Spinal Cord Injury as AIS A (sensory and motor complete) and B (sensory incomplete and motor complete), and eight age-matched controls (mean age=31.6±2.2 years, 5 female, $p > 0.05$, $\mathrm{Wilcoxon\ rank-sum\ test}$) participated in the study conducted at The Miami Project to Cure Paralysis, University of Miami (see clinical information of individuals with tetraplegia in **Supplementary Table 1**). All experiments were approved by the Ethical Review Board of University of Miami in compliance with the Declaration of Helsinki; all participants provided informed consent to participate in the study. Each participant engaged in a single experimental session which lasted about three hours. First, participants were familiarized with ChinMotion during a brief calibration period which lasted about thirty minutes (ChinMotion calibration is illustrated in **Supplementary Video 3**). After calibrating ChinMotion, each uninjured participant performed three experimental tasks in the following order: (1) pointing using ChinMotion (see evaluation of pointing performance in Methods; **Supplementary Video 1**), (2) 3D control of a virtual robotic arm using ChinMotion (see evaluation of 3D control of a virtual robotic arm in Methods; **Supplementary Video 2**), and (3) pointing using a keyboard-mounted pointing stick (**Fig. 1b**) with one hand. Individuals with tetraplegia performed only (1) and (2).

**ChinMotion.** A triple axis accelerometer (InvenSense MPU 6050) and a conductive rubber stretchable cord (Adafruit) transduce deviations from free fall and stretching forces induced by movements of chin, lips and tongue into electrical signals. These are processed by an Arduino Leonardo clone microcontroller (Polulu A-Star 32U4 Micro) and passed to a host computer via USB interface (**Fig.1a**). To control cursor movements on the host computer, the microcontroller was programmed using the open-source Arduino Software IDE with the code available at **Supplementary Software 1**. In short, this code translates accelerometer and cord measurements surpassing a set of predefined thresholds into XY cursor displacements and click function (pointing; **Fig. 2a**) or Z displacement (robotic arm; **Fig. 2d, e**). To calibrate ChinMotion, a graphical user interface (**Supplementary Software 2**) written using the open-source Processing 2 software allows for the interactive customization of thresholds and cursor speeds (ChinMotion calibration is illustrated in **Supplementary Video 3**) that ensured the establishment of intuitive kinaesthetic mappings between user motions and pointer function. **Supplementary Step-by-Step Instructions** contains extended information (parts, stepwise instructions, and schematics) required for replicating ChinMotion.

**Evaluation of point-and-click performance.** To evaluate point-and-click performance, we applied Fitt's law to a centre-out-centre reach-and-click task adapting the ISO9241-9 standard recommendations[19]. To complete the task, participants moved the pointer from the centre of the screen to select a randomly-appearing peripheral target and then moved back to select the centre (**Supplementary Video 1**). Participants were instructed to move towards the target and click within it as fast as possible. A trial ended successfully when the click occurred within the target. If a click occurred elsewhere, it stopped the pointer and penalized performance by increasing the effective target selection time. Trials with clicks occurring outside the target were further registered to estimate error rates. There were 50 possible targets including the centre (**Fig. 2a**), with 8 different orientations (0, 45, 90, 135, 180, 225, 270, and 315 degrees), 2 different widths (30 and 61 pixels), and 3 different distances from the centre (122, 244, and 300 pixels). Following Shanon's formulation[20], such an arrangement allowed the evaluation of $m = 6$ index of difficulty ($ID$) conditions (**Supplementary Table 7**). To account for discrepancies between participant's effective performance and experimental $ID$ conditions; that is, the different spread of performed movement end-points across conditions, we adjusted for accuracy per each participant[19]:

$$ID_e = log_2\left(\frac{De}{4.133 \times SD} + 1\right). \qquad (1)$$

The effective distance ($De$) represents the mean movement distance between measured starting and end positions, and $SD$ is the standard deviation of $w$, which is the distance between measured end positions and the centre of the target. Note that assuming that $w$ is normally distributed, its entropy $H(w)$ is a function of $SD$:

$H(w) = B \times log_2\left(\sqrt{2\pi e} \times SD\right) = B \times log_2\,(4.133 \times SD)$, where $B$ is the bandwidth of the communication channel[19]. $ID_e$ is therefore measured in bits. Considering $n$ = 8 number of participants, this resulted in $N = 48$ ($N = n \times m$) pairs of effective target selection times and adjusted index of difficulty data ($STe_{ij}, IDe_{ij}$) where $STe_{ij}$ represents the mean trial selection time and $1 \leq i \leq n, 1 \leq j \leq m$. Note that in our task, target selection times encompass reaction, movement, and click times. Least-squares linear regression was used to find the intercept $a$ and the slope $b$ parameters of Fitt's law:

$$ST = a + bIDe, \qquad (2)$$

which linearly relates target selection time ($ST$) and effective index of difficulty ($ID_e$). Regression models for participants with tetraplegia using ChinMotion and uninjured participants using both ChinMotion and pointing stick all confirmed Fitt's law goodness of fit (**Supplementary Tables 2, 5** and **Supplementary Figure 1**), enabling the use of throughput ($TP$) as a pointing performance metric:

$$TP = \frac{1}{n}\sum_{i}^{n}\left(\frac{1}{m}\sum_{j=1}^{m}\frac{IDe_{ij}}{STe_{ij}}\right), \qquad (3)$$

$TP$ is a combined measure of speed and accuracy which is measured in bits per second (bits/s). In addition, error rates were estimated as the percentage (%) of trials with clicks occurring outside the target. Each participant completed two runs of 50 trials (100 trials in total). This task was programmed (**Supplementary Software 3**) using the open-source Processing 2 software.

Within the ISO9241-9 framework, our centre-out-centre reach-and-click task allowed for comparison with mainstream hand-operated pointing devices[19] (**Supplementary Table 3**), and hands-free body- and brain- machine interfaces with the highest performance reported to date evaluating the same metrics: the Tongue Drive System (TDS)[4] (**Supplementary Table 4**) and the BrainGate neuroprosthetic system[11]. However, the evaluation of the BrainGate pilot clinical trial[11] excluded all trials with selection times greater than 25 seconds which can alter the estimation of Fitt's law parameters. This challenges the comparability of the results to studies that did not exclude trials on that basis. To illustrate such effects and to enable an unbiased comparison between ChinMotion and BrainGate systems we estimated additional

Fitt's law regression models excluding trials fulfilling the same criterion (**Supplementary Fig. 1b; Supplementary Table 5**). The BrainGate pilot clinical trial[11] used Fitt's law parameters as performance metric instead of throughput; **Supplementary Table 6** compares ChinMotion and BrainGate pointing performance using Fitt's law parameters, error rates (false click rates in the BrainGate pilot clinical trial[11]), and percentage of excluded trials (time out errors in the BrainGate pilot clinical trial[11]).

Although other hands-free interfaces have been evaluated following the ISO9241-9 standard: Sip-&-Puff[4], head orientation and electromyography based[8] body-machine interfaces, and the BrainGate2 neuroprosthetic system[13]; their performance is not comparable as they only enabled 2D pointing without active click control. In these studies click function was virtually enabled with arbitrary target dwelled times ranging from 0.5 to 2 seconds which lessens task complexity and interface usability.

**Evaluation of 3D control of a virtual robotic arm.** We evaluated 3D control of a virtual three degrees of freedom simulation of FabLab RUC Arm which is freely available for download online (http://fablab.ruc.dk/robot-arm-v-0-1/). Participants used ChinMotion to control the arm endpoint position in a 3D Cartesian space ($1m^3$; **Fig. 2d, e**) without computer assistance. Participants engaged in a centre-out-centre reach-and-hold task (**Fig. 2d, e; Supplementary Video 2; Supplementary Software 4**) programmed using the open-source Processing 2 software. A trial consisted in moving the arm endpoint from a starting location (red sphere of 5 cm radius in **Fig. 2d**) to hold within the volume of a randomly selected target (of a total of 18 yellow spheres of 5cm radius in **Fig. 2d, e**) for 1 s. The arm endpoint was then returned to within the starting location volume for a further 1 s. Participants were instructed to complete a total of 20 trials as fast as possible. Average trial completion time was used as performance metric. Trial 1 was excluded from the analysis as it was allocated for practice. Participants performed the task in first person view, with only the end effector of the arm visible, and with additional visual cues facilitating 3D position estimation within the rendered task layout (**Fig. 2e, Supplementary Video 2**).

**Statistical analysis.** Results are reported within the text as mean ± standard error of the mean (SEM). Within **Fig. 2**, the central line in box-plots is the median, the box is defined by 25th and 75th percentiles, the whiskers extend to the most extreme data points not considered outliers, and outliers are plotted individually (~ ±2.7σ). Comparisons between individuals with tetraplegia and uninjured controls were examined using two-sided Wilcoxon rank-sum test (**Fig. 2b-c, f**). Throughput and error rate comparisons between uninjured controls using ChinMotion and pointing stick were examined using two-sided Wilcoxon signed rank test (**Fig. 2b-c**). The sample size was not chosen with consideration of adequate power to detect a prespecified effect. Regression models for Fitt's law were tested against a constant model using the F-test (**Supplementary Tables 2 and 5; Supplementary Fig. 1**). The data meet all the assumptions of the tests used. All analysis was performed using the MATLAB environment (The MathsWorks, Natick, MA).

## Acknowledgements

The authors would like to thank all the participants, Felipe de Carvalho for helping in designing ChinMotion board and helping in preparing Supplementary Schematics, Robert Camarena for helping in preparing Supplementary Videos and Supplementary Step-by-Step Instructions, Demetris Soteropoulos for helping in revising Supplementary Software and Supplementary Step-by-Step Instructions, and Sarah Lehman for administrative support. The study was supported by the Wellcome Trust and the National Institutes of Health.

## Author contributions

F.G. conceived and developed ChinMotion, designed the experiments and analysed the data. M.A.P. and S.N.B. supervised all aspects of the work. F.G., S.N.B. and M.A.P. wrote the paper. M.A.P. was responsible for research infrastructure and governance.

## Competing financial interests

No conflicts of interest, financial or otherwise, are declared by the authors.

## Materials and Correspondence

Correspondence should be addressed to F.G. (ferran.galan@ncl.ac.uk)

# Figure Legends

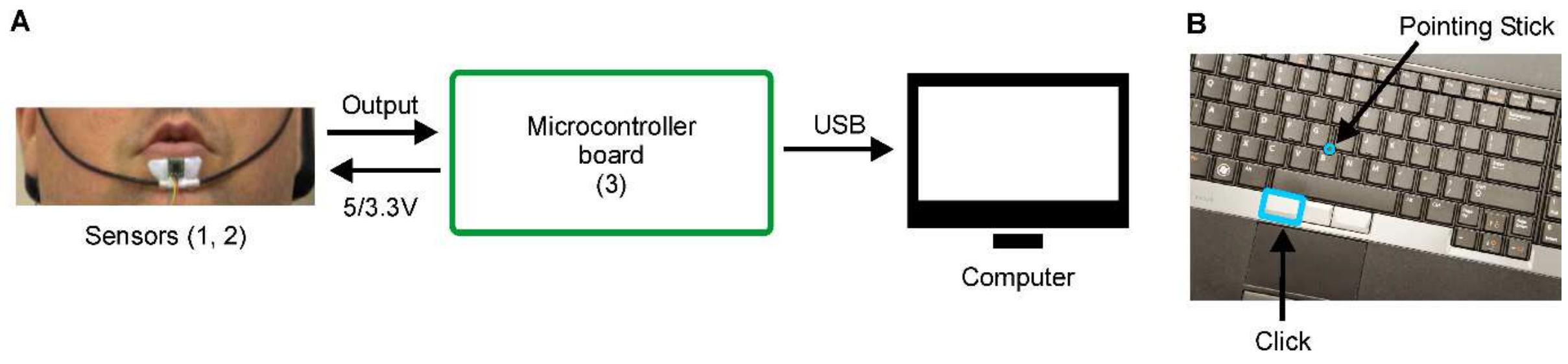

**Figure 1. ChinMotion and keyboard mounted pointing stick. (a)** System-level diagram of ChinMotion and its three main components: (1) a triple axis accelerometer placed below the lower lip, (2) a conductive rubber stretchable cord sensor following the contour of the face, and (3) a microcontroller board. **(b)** Uninjured participants were instructed to use a keyboard mounted pointing stick and click button with one hand.

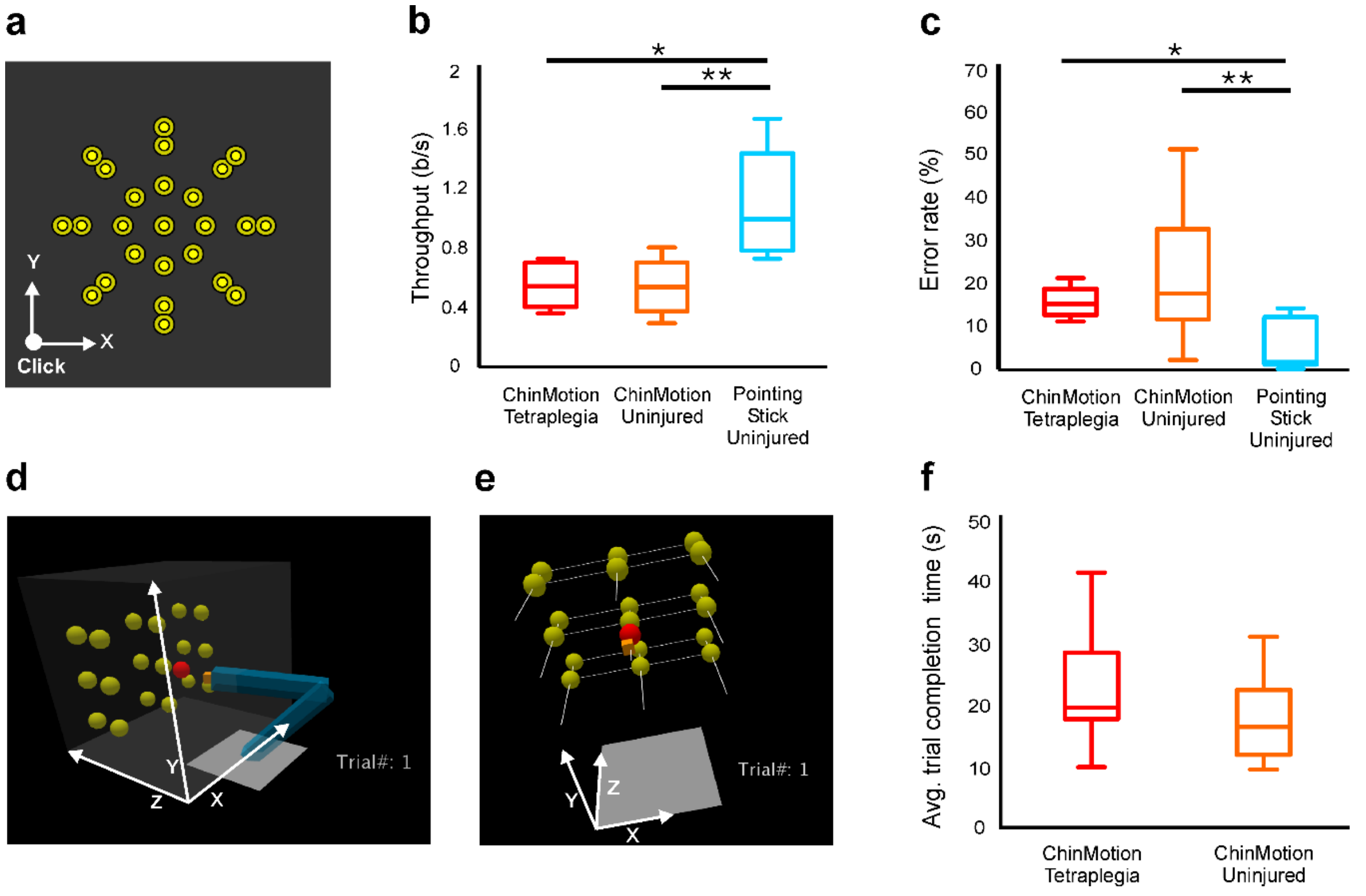

**Figure 2. Experimental tasks and performance metrics. (a)** Task layout utilized for evaluating point-and-click performance (see Methods). There were a total of 50 possible targets; during task execution only one target per trial was visible. ChinMotion enabled participants (1) to move the computer pointer in XY coordinates by producing lateral, retrusive, and protrusive movements of chin and lips, and (2) to click via protrusive movements of the tongue (see **Supplementary Video 1**). **(b)** Throughput (see Methods) achieved by individuals with tetraplegia using ChinMotion (*ChinMotion Tetraplegia;* n = 8), by uninjured controls using ChinMotion (*ChinMotion Uninjured*, n = 8), and by uninjured controls using a keyboard mounted pointing stick (*Pointing Stick Uninjured;* n = 8). $*p < 0.001$, Wilcoxon rank − sum test; $**\ p < 0.01$, Wilcoxon signed rank test. **(c)** Error rates (see Methods) reached by *ChinMotion Tetraplegia* (n = 8), *ChinMotion Uninjured* (n = 8), and *Pointing Stick Uninjured* (n = 8). $*p < 0.01$, Wilcoxon rank − sum test; $**p < 0.05$, Wilcoxon signed rank test. **(d)** Task layout utilized for evaluating 3D control of a virtual robotic arm (see Methods). ChinMotion enabled participants (1) to move the robotic

arm in XY coordinates by producing lateral, retrusive, and protrusive movements of chin and lips, and (2) to move it in Z coordinates via protrusive movements of the tongue (see **Supplementary Video 2**). **(e)** Participants performed the task in first person view, with only the effector of the arm visible (orange block), and additional visual cues (white lines) facilitating 3D position estimation within the rendered task layout (see **Supplementary Video 2**). **(f)** Average trial completion time (see Methods) achieved by *ChinMotion Tetraplegia* (n = 8) and *ChinMotion Uninjured* (n = 8).

# Supplementary Information

Ferran Galán[1,2*], Stuart N. Baker[2] and Monica A. Perez[1]

[1]Department of Neurological Surgery, The Miami Project to Cure Paralysis, University of Miami. [2]Institute of Neuroscience, Newcastle University.

Correspondence to:
Ferran Galán P.hD.
Henry Wellcome Building
The Medical School, Framlington Place
Newcastle upon Tyne, NE2 4HH, UK
Telephone: +44 (0) 191 208 3271
Fax: +44(0) 191 208 5227
ferran.galan@ncl.ac.uk

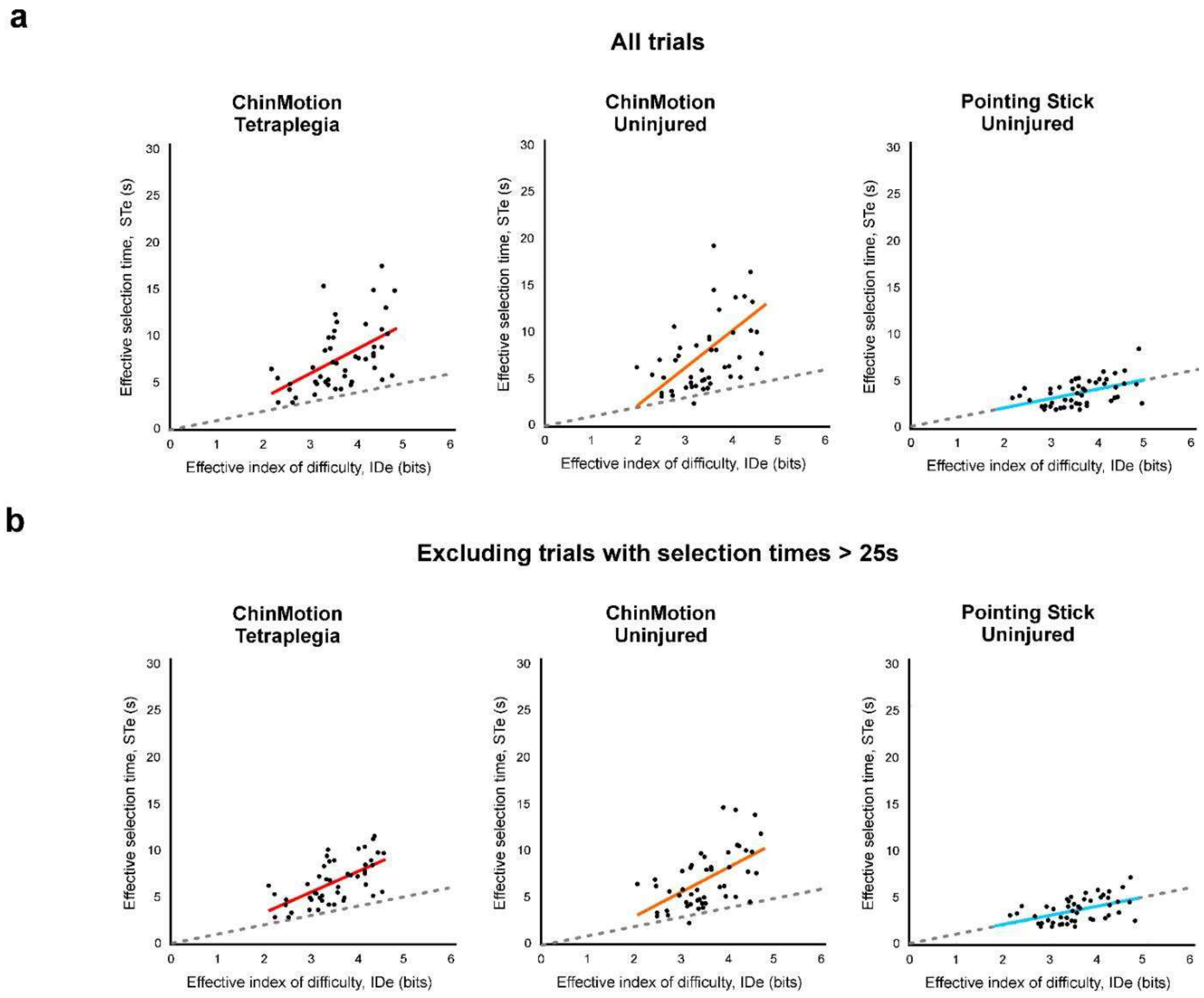

**Supplementary Figure 1. Visualization of participants' point-and-click performance. (a)** Scatter plots of $N = 8 \times 6$ pairs of effective target selection times and effective index of difficulty data. Lines in each plot are the least-squares Fitt's law regression models, which are summarized in **Supplementary Table 2,** of all points shown. Grey dashed lines represent $ST_e = ID_e$. **(b)** Scatter plots of $N = 8 \times 6$ pairs of effective target selection times and effective index of difficulty data after excluding trials with target selection times greater than 25 seconds. See Methods. Lines in each plot are the least-squares Fitt's law regression models, which are summarized in **Supplementary Table 5,** of all points shown. Note that the exclusion of such trials (representing 2.1± 1.0, 4.9 ± 2.9 and 0.1 ± 0.1 % of the total number of trials performed by *ChinMotion Tetraplegia*, *ChinMotion Uninjured*, and *Pointing Stick*

*Uninjured* respectively) specially reduces *ChinMotion Tetraplegia* and *ChinMotion Uninjured* effective target selection times ($STe$), bringing Fitt's intercepts closer to 0, and reducing Fitt's slope.

| Participant | Sex | Age (years) | SCI level | Post injury (months) | AIS | Medication | Spasm score |
|---|---|---|---|---|---|---|---|
| 1 | M | 43 | C3 | 276 | B | Baclofen | 4 |
| 2 | F | 38 | C4 | 144 | A | Baclofen | 4 |
| 3 | F | 25 | C2 | 216 | A | Clonopin | 4 |
| 4 | M | 34 | C4 | 216 | A | Baclofen/Ditropen | 3 |
| 5 | M | 40 | C2 | 62 | A | None | 3 |
| 6 | M | 32 | C4 | 168 | B | Valium | 4 |
| 7 | M | 38 | C2 | 127 | A | None | 4 |
| 8 | M | 44 | C3 | 97 | B | Baclofen | 4 |

**Supplementary Table 1. Demographic and clinical details of participants with tetraplegia due to SCI.** M: male; F: female; AIS, American Spinal Injury Association Impairment Scale; A: sensory and motor complete; B: sensory incomplete and motor complete. Spasm score; 0: no spasms, 1: one or fewer spasms per day, 2: between 1 and 5 spasms per day, 3: 5 to <10 spasms per day, and 4: 10 or more spasms per day.

| | *ChinMotion Tetraplegia* | *ChinMotion Uninjured* | *Pointing Stick Uninjured* |
|---|---|---|---|
| **Intercept, a (s)** | -1.73±2.23 | -5.59±3.74 | 0.02±0.95 |
| **Slope, b (s/bits)** | 2.61±0.62 | 3.95 ±1.07 | 1.01±0.26 |
| **R-squared** | 0.28 | 0.23 | 0.24 |
| **F-statistic vs. constant model** | 17.5 | 13.7 | 14.4 |
| **N** | 48 | 48 | 48 |
| **d.f** | 46 | 46 | 46 |
| **p-value** | <0.0005 | <0.001 | <0.001 |
| **Throughput, TP (bits/s)** | 0.54±0.05 | 0.55 ± 0.07 | 1.10±0.13 |
| **Error Rate (%)** | 18.38±3.8 | 22.00±5.6 | 5.38±2.2 |

**Supplementary Table 2. Fitt's law regression models and measures of performance.**

Fitt's law regression models were estimated via least-squares from $N = 8 \times 6$ pairs of effective target selection times and adjusted index of difficulty data (see Methods) depicted in **Supplementary Fig. 1a**.

| Device | Throughput, TP (bits/s) (higher is better) |
|---|---|
| Isometric joystick[19] | 1.6-2.55 |
| Touchpad[19] | 0.99-2.9 |
| Mouse[19] | 3.7-4.9 |
| *Pointing Stick Uninjured** | 1.10±0.13 |
| *ChinMotion Tetraplegia** | 0.54±0.05 |
| *ChinMotion Uninjured** | 0.55±0.07 |

**Supplementary Table 3. ChinMotion compared to mainstream hand-operated pointing devices[19].** *Values from **Supplementary Table 2**.

| | **Throughput, TP (bits/s)** (higher is better) | **Error Rate (%)** (lower is better) |
|---|---|---|
| ***ChinMotion Tetraplegia**** **(within 1 session )** | 0.54±0.05 | 18.38±3.8 |
| ***ChinMotion Uninjured**** **(within 1 session)** | 0.55±0.07 | 22.00±5.6 |
| **TDS Tetraplegia[4] (trial day 1)** | 0.29±0.21 | 64.68±25.2 |
| **TDS Tetraplegia[4] (trial day 6)** | 0.72±0.41 | 41.48±26.0 |

**Supplementary Table 4. ChinMotion compared to Tongue Drive System[4] (TDS).** See Methods.*Values from **Supplementary Tables 2-3.**

| | *ChinMotion Tetraplegia* | *ChinMotion Uninjured* | *Pointing Stick Uninjured* |
|---|---|---|---|
| **Intercept, *a* (s)** | -1.09±1.42 | -2.24±1.99 | 0.14±0.93 |
| **Slope, *b* (s/bits)** | 2.20±0.40 | 2.63±0.56 | 0.97±0.26 |
| **R-squared** | 0.4 | 0.32 | 0.24 |
| **F-statistic vs. constant model** | 30.6 | 22.2 | 14.5 |
| **N** | 48 | 48 | 48 |
| **d.f** | 46 | 46 | 46 |
| **p-value** | $<10^{-5}$ | $<10^{-5}$ | <0.0005 |
| **Throughput, TP (bits/s)** | 0.58±0.05 | 0.58±0.06 | 1.09±0.11 |
| **Error Rate (%)** | 18.38±3.8 | 24.63±5.4 | 5.38±2.2 |

**Supplementary Table 5. Fitt's law regression models and measures of performance excluding trials with target selection times greater than 25 seconds (time out errors) as in the BrainGait pilot clinical trial[11].** See Methods. Such trials represented 2.1± 1.0, 4.9 ± 2.9 and 0.1 ± 0.1 % of the total number of trials performed by *ChinMotion Tetraplegia*, *ChinMotion Uninjured*, and *Pointing Stick Uninjured* respectively. Fitt's law regression models were estimated via least-squares from $N = 8 \times 6$ pairs of effective target selection times and adjusted index of difficulty data (see Methods) depicted in **Supplementary Fig. 1b**. Note that excluding trials with target selection times greater than 25 seconds specially altered *ChinMotion Tetraplegia* and *ChinMotion Uninjured* Fitt's law regression models and measures of performance shown in **Supplementary Table 2** by bringing Fitt's intercepts closer to 0, reducing Fitt's slope, and increasing throughput. Such alterations are further illustrated in **Supplementary Fig. 1**.

| | ***ChinMotion Tetraplegia*** ***** **(within 1 session)** | ***ChinMotion Uninjured*** ***** **(within 1 session)** | **BrainGate S3 (Averaging trial days 999–1003)** |
|---|---|---|---|
| **Fitt's Intercept, *a* (s)** (zero is better) | -1.09±1.42 | -2.24±1.99 | 0.8 |
| **Fitt's Slope, *b* (s/bits)** (lower is better) | 2.20±0.40 | 2.63±0.56 | 3.3 |
| **Error Rates / FCR**[11] (%; lower is better) | 18.38±3.8 | 24.63±5.4 | 41 |
| **Excluded Trials / Time Out Errors**[11] (%; lower is better) | 2.1±1.0 | 4.9±2.9 | 8.1 |

**Supplementary Table 6. ChinMotion compared to BrainGate neuroprosthetic system[11].**

See Methods.*Fitt's parameters from **Supplementary Table 5**. FCR: false click rates.

| ID (bits) | | Widths (pixels) 30 | 61 |
|---|---|---|---|
| Distances (pixels) | 122 | 2.34 | 1.58 |
| | 244 | 3.19 | 2.32 |
| | 300 | 3.46 | 2.57 |

**Supplementary Table 7. Index of difficulty (ID) conditions.**

**Supplementary Software 1. *ChinMotion.ino* sketch.** When uploaded to Polulu A-Star 32U4 Micro, this code translates accelerometer and cord measurements surpassing a set of predefined thresholds into XY cursor displacements and click function (pointing-and clicking; **fig. 2a**) or Z displacement (robotic arm; **fig. 2d, e**). It was written and uploaded to Polulu A-Star 32U4 Micro using Arduino IDE software freely available at https://www.arduino.cc/en/Main/Software.

**Supplementary Software 2. *Calibration.pde* sketch.** Graphical user interface which allows for the interactive customization of thresholds and cursor speeds that determine ChinMotion function (ChinMotion calibration is illustrated in **Supplementary Video 3**). Can be edited and executed using Processing 2 software freely available at https://processing.org/.

**Supplementary Software 3. *Pointing.pde* sketch.** Program implementing the center-out-center reach-and-click task (see **Fig. 2a**) utilized in this study. Can be edited and executed using Processing 2 software freely available at https://processing.org/.

**Supplementary Software 4. *RobotArm.pde* sketch.** Program implementing the center-out-center reach-and-hold task (see **Fig. 2d, e**) utilized in this study. Can be edited and executed using Processing 2 software freely available at https://processing.org/.

**Supplementary Schematics. *ChinMotion_BreakOutBoard_v1.sch*** and ***ChinMotion_BreakOutBoard_v1.brd* eagle files.** ChinMotion board schematics.

**Supplementary Video 1.** Video illustrating point-and-click using ChinMotion by a participant with tetraplegia (Participant 7 in **Supplementary Table 1**; see **Fig. 2a; Supplementary Software 3**).

**Supplementary Video 2.** Video illustrating 3D control of a virtual robotic arm using ChinMotion by a participant with high cervical tetraplegia (Participant 7 in **Supplementary Table 1**; see **Fig. 2d,e; Supplementary Software 4)**.

**Supplementary Video 3.** Video explaining how to calibrate ChinMotion using **Supplementary Software 2**.

Supplementary videos were recorded in one session approximately two months after the completion of the study for demonstration purposes. Participant 7 provided informed consent to publish photos and videos taken during this recording session. Participant 7 did not have any contact with ChinMotion during the period in between experimental and recording sessions. Note that the picture-in-picture videos were flipped horizontally for illustration purposes.

## Supplementary Step-by-Step Instructions

Our goal with ChinMotion is to stimulate the innovation of affordable, fast to fabricate, easily reproducible and customizable interfaces capable of improving the quality of life to the greatest number of individuals with paralysis. ChinMotion can be built in a couple of hours utilizing off-the shelf parts, open-source hardware and freely available software for less than 40 USD per unit. This section enumerates the parts and describes the build process allowing the replication of ChinMotion as utilized in our study; we expect that it can be of import to a general audience interested in using ChinMotion for educational, research and development purposes. ChinMotion's hardware builds upon and modifies Head Mouse by Martin Millmore which is publically available at: http://www.instructables.com/id/Head-Mouse-Game-controller-or-disability-aid/.

**Parts list**

1. 1 × Pololu A-Star 32U4 Micro; the controller is a tiny Arduino Leonardo clone.
2. 1 × USB A to micro B Cable connects Pololu A-Star 32U4 Micro with the host computer.
3. 1 × ChinMotion board is a minimized (11 × 11 mm) version of SparkFun Triple Axis Accelerometer and Gyro Breakout - MPU-6050 which facilitates its attachment to the chin. Eagle files can be found at **Supplementary Schematics 1**.
4. 1 × Conductive Rubber Cord Stretch Sensor.
5. 1 × A 3.3V to 5V logic level converter; while Pololu A-Star 32U4 Micro works at 5V, ChinMotion board works at 3.3V.
6. 2 × 10K resistors.
7. 1 × Push button toggles the mouse control state between ChinMotion and computer mouse.

8. Silicone wires; connections to ChinMotion board need to be flexible introducing minimal tension.
9. Velcro allows the attachment of ChinMotion board + Conductive Rubber Cord Stretch Sensor to the adhesive electrode (see below).
10. Adhesive pad facilitate the attachment of ChinMotion board + Conductive Rubber Cord Stretch Sensor to the chin. We have been making pads from H91SG: ECG electrode, foam, oval, 50/pkg; each electrode allows for two uses. Any other skin-friendly adhesive may be suitable.
11. 1 × Enclosure box fits in the Pololu A-Star 32U4 Micro, the A 3.3V to 5V logic level converter, push button and connections.
12. 2 × croc clips.
13. Arduino IDE. Installing Arduino IDE in the host computer is required for running **Supplementary Software 1; *ChinMotion.ino* sketch.**
14. Processing 2. Installing Processing 2 in the host computer is required for running **Supplementary Softwares 2-4; *Calibration.pde, Pointing.pde, RobotArm.pde*.**

See *ChinMotion* parts in **Illustration 1.**

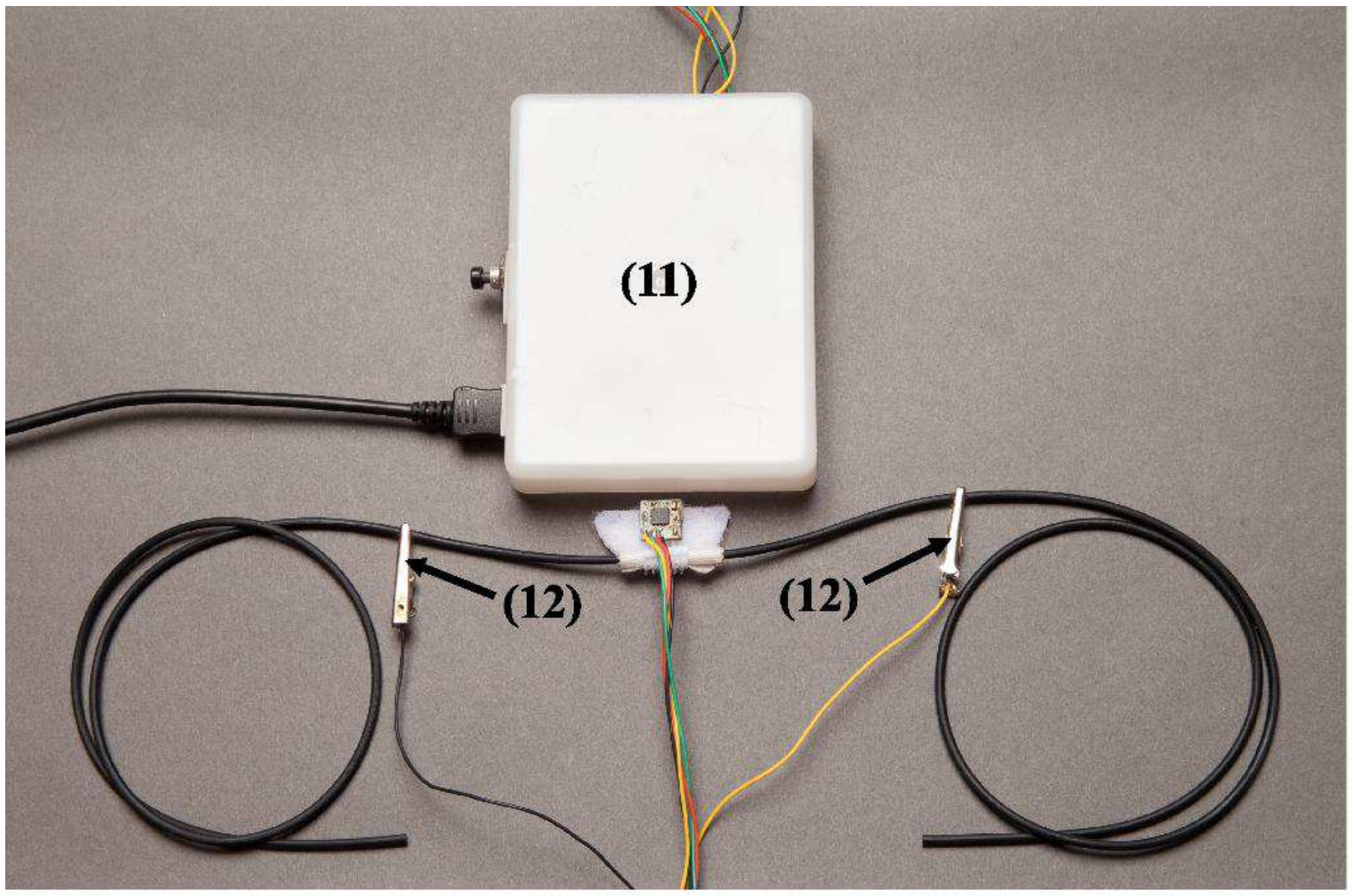

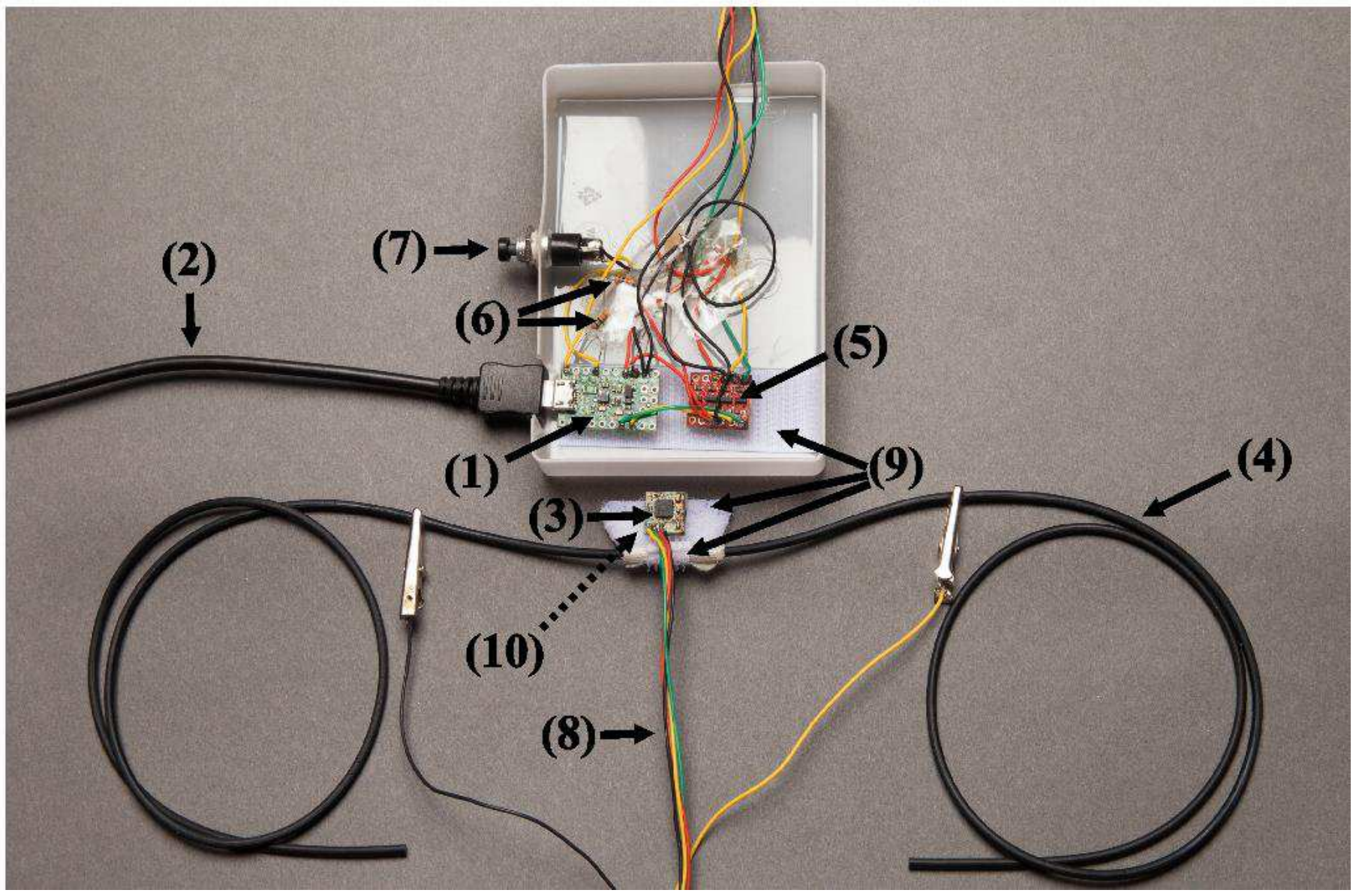

**Illustration 1.** ***ChinMotion*** **parts.** Note that the adhesive pad (10) is not visible as it is occluded by Velcro (9); see step 5 in build process.

**Build process**

1. Make three holes in the enclosure box (11). The first will hold the push button (7), the second will allow the silicone wires connecting ChinMotion board (3) and Conductive Rubber Cord Stretch Sensor (4) to leave the box, the third will allow connection of the USB A to micro B Cable (2) to Pololu A-Star 32U4 Micro (1). **See Illustration 1**.
2. Fit the push button (7) and secure Pololu A-Star 32U4 Micro (1) and 3.3V to 5V logic level converter (5) in the enclosure box (11). In **Illustration 1** they are secured using Velcro (9).
3. Wire all up as illustrated below; the silicone wires (8) connecting ChinMotion board (3) and Conductive Rubber Cord Stretch Sensor (4) should be long enough to reach the user's chin.

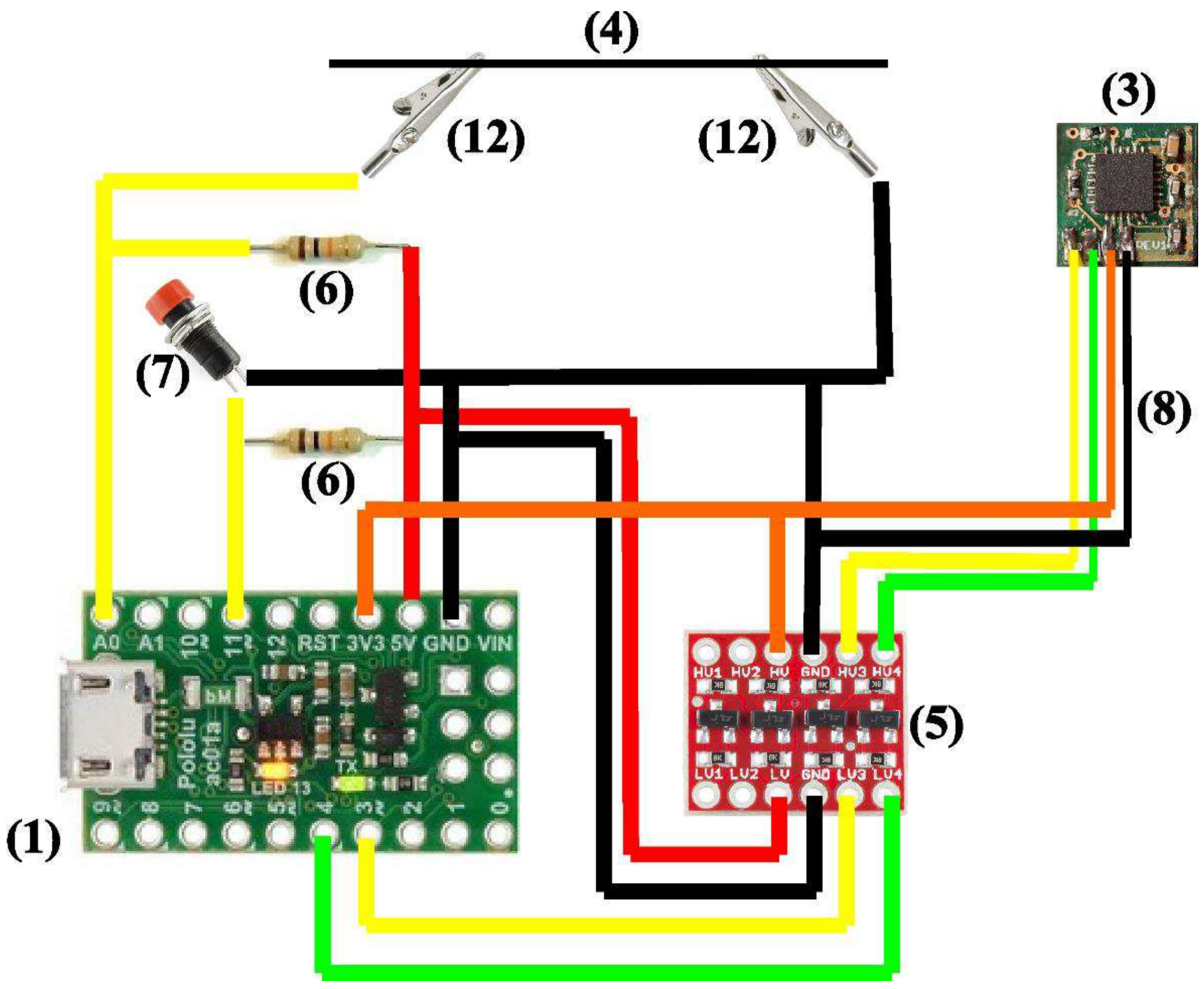

4. Cut an adhesive pad (10) as shown:

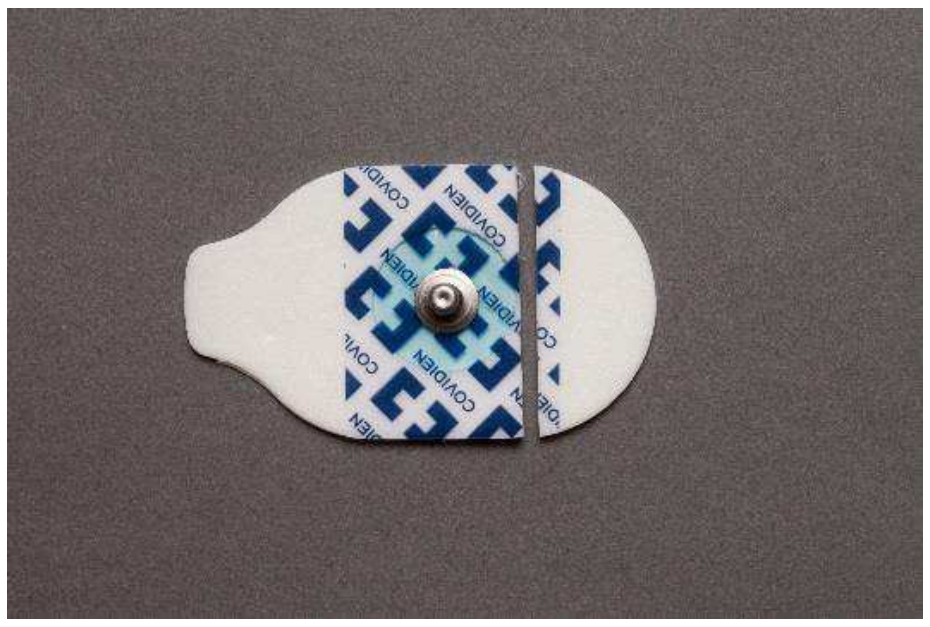

5. Cut Velcro (9; soft side) and fix it on the pad (10):

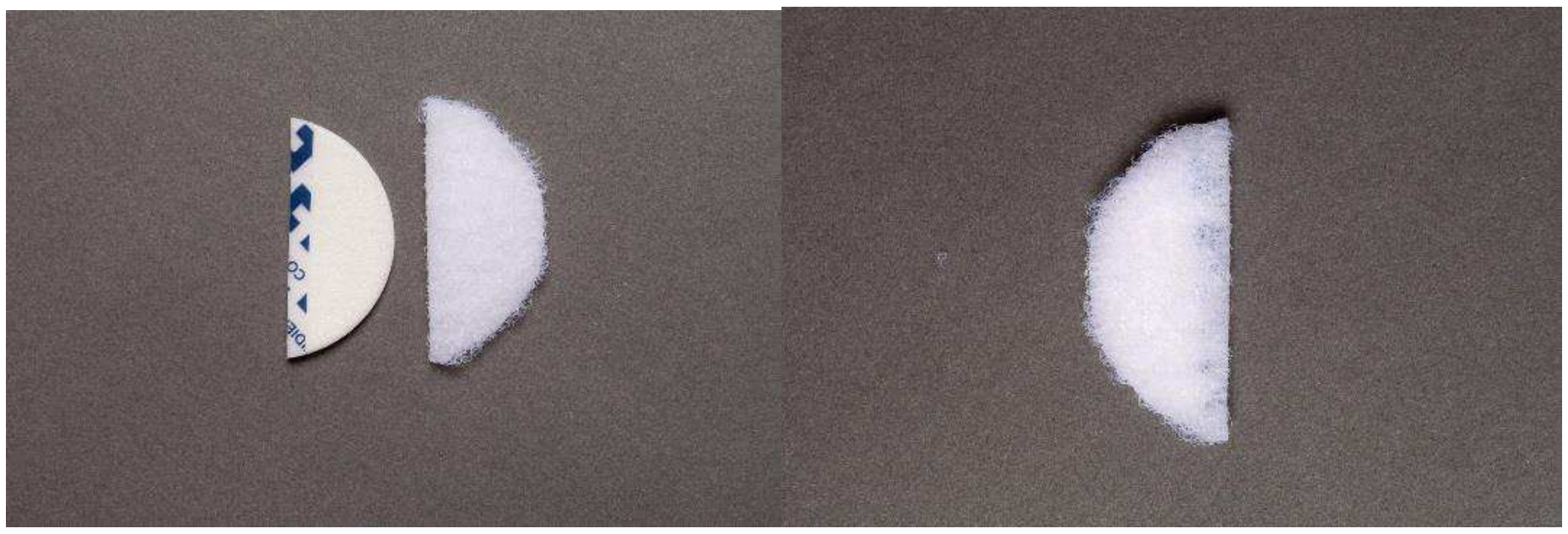

6. Cut Velcro (9; rough side) and fix it at the back of ChinMotion board (3):

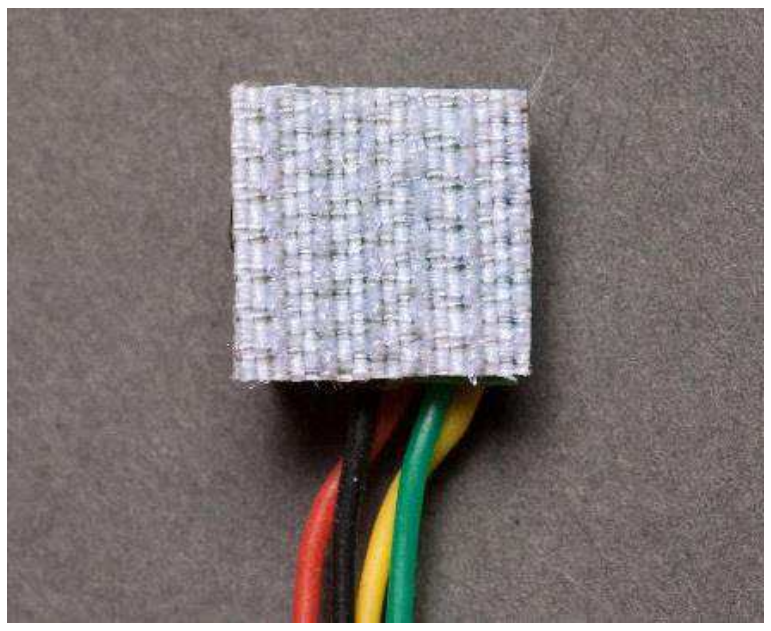

7. Cut Velcro (9; rough side) and fix it around the Conductive Rubber Cord Stretch Sensor (4) at the middle point; building volume around the cord with tape may be useful:

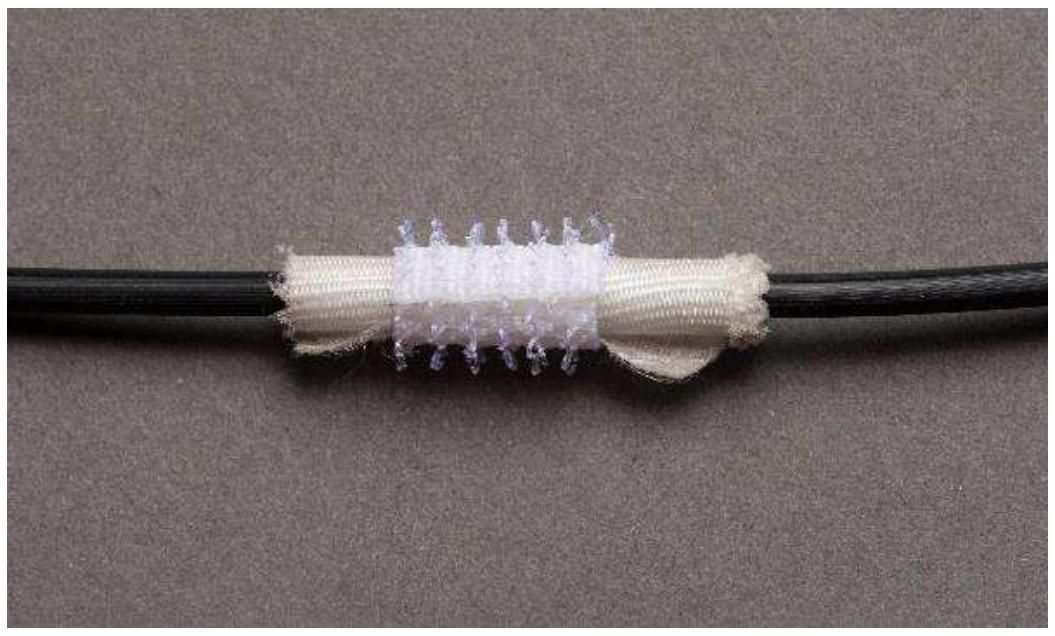

8. Put together ChinMotion board (3), Conductive Rubber Cord Stretch Sensor (4), and adhesive pad (10):

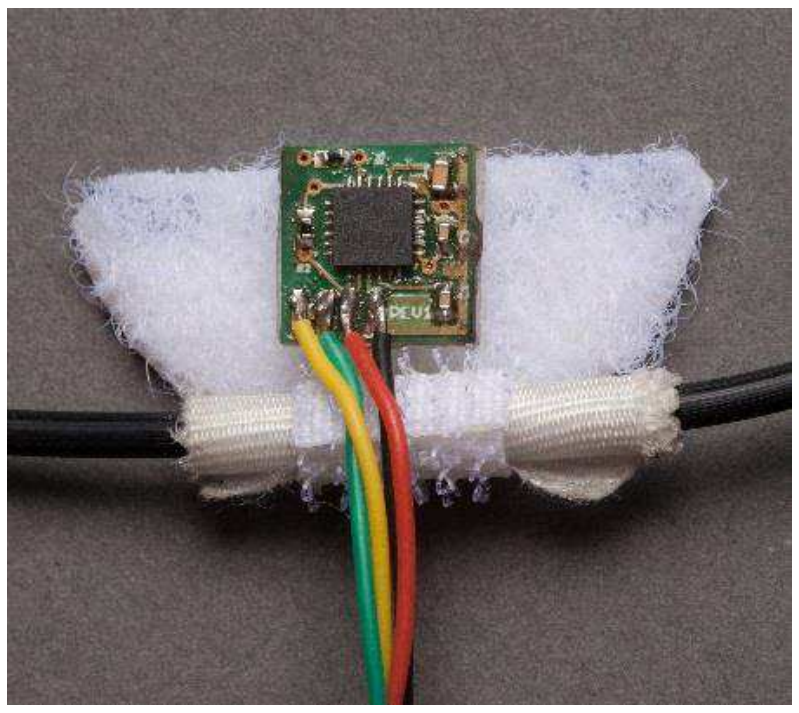

**Setup process**

1. Install Arduino IDE software freely available at https://www.arduino.cc/en/Main/Software.
2. If you are using a host computer running Microsoft Windows you must install Pololu A-Star 32U4 Micro drivers as described at https://www.pololu.com/docs/0J61/6.21.
3. Setup Pololu A-Star 32U4 Micro as described at https://www.pololu.com/docs/0J61/6.2.
4. Upload **Supplementary Software 1 (*ChinMotion.ino* sketch)** to Pololu A-Star 32U4 Micro.
5. Install Processing 2 software (version 2.2.1) freely available at https://processing.org/.
6. Go to Sketch, Import Library, Add Library, and search for: Arduino (Firmdata), ControlP5, OCD (Obssesive Camera Direction), PeasyCam, and Signal Filter. Install them all; they are libraries utilised by **Supplementary Softwares 2-4 (*Calibration.pde, Pointing.pde*, *RobotArm.pde*)**.
7. Stick ChinMotion sensors (on the pad as in step 8 of build process) on the user's chin as in **Fig. 1** and **Supplementary Videos**. Route the Conductive Rubber Cord Stretch Sensor (4) following the contour of the face until it sits on the user's ears.

8. Secure both ends of Conductive Rubber Cord Stretch Sensor (4) on the user's neck with skin-friendly tape. Adjust the tension while calibrating to ensure that protrusive tongue movements against the Conductive Rubber Cord Stretch Sensor (4) are salient but do not interfere with ChinMotion board's (3) readings (see next step).
9. Open **Supplementary Software 2 (*Calibration.pde* sketch)** and set the path (line 24) to save the *SaveThresholds.txt* file with thresholds and mouse speed in the same folder with **Supplementary Software 4 (*RobotArm.pde* sketch)**. Edit also line 25 specifying the port used by Pololu A-Star 32U4 Micro.
10. Run **Supplementary Software 2 (*Calibration.pde* sketch)** and calibrate ChinMotion as illustrated in **Supplementary Video 3**. If any of the traces is not visible; it may be out of range. Try to adjust ChinMotion board's position (3) and/or Conductive Rubber Cord Stretch Sensor's (4) tension.
11. Once calibrated, do not kill ***Calibration.pde*** sketch. Switch on ChinMotion with the push button and it will take over the computer mouse. Switch off ChinMotion with the push button get back the computer mouse. ***Calibration.pde*** sketch must be running.

***Note 1:*** Before running **Supplementary Software 3 (*Pointing.pde* sketch)** and **Supplementary Software 4 (*RobotArm.pde* sketch),** edit lines 27 and 34 respectively which specify the path where to save the *output.txt* file.

***Note 2:*** Before running ***RobotArm.pde***, edit line 35 specifying the port used by Pololu A-Star 32U4 Micro; **Supplementary Software 2 (*Calibration.pde* sketch)** must be also closed. You may also need to edit line 185 to adjust a suitable Y-mapping; note that YZ axis in ***RobotArm.pde*** are exchanged with respect **Figure 2d,e**.

***Note 3:*** Processing 2 software and ***RobotArm.pde*** use OpenGL for 3D graphics which may cause problems on old hardware and/or out-of-date drivers. Check the following documentation for issues: https://github.com/processing/processing/wiki/OpenGL-Issues